# Chaotic electro-convection flow states of a dielectric liquid between two parallel electrodes

Qi Wang[1, 2], Yifei Guan[3], Junyu Huang[1, 2], Jian Wu[1, 2*]

[1] *School of Energy Science and Engineering, Harbin Institute of Technology, Harbin 150001, PR China*
[2] *Key Laboratory of Aerospace Thermophysics, Ministry of Industry and Information Technology, Harbin 150001, PR China*
[3] *Department of Mechanical Engineering, Rice University, Houston, TX, U.S.A. 77005*

Apr 2021

**Abstract**

The two-dimensional regular and chaotic electro-convective flow states of a dielectric liquid between two infinite parallel planar electrodes are investigated using a two-relaxation-time lattice Boltzmann method. Positive charges injected at the metallic planar electrode located at the bottom of the dielectric liquid layer are transported towards the grounded upper electrode by the synergy of the flow and the electric field. The various flow states can be characterized by a non-dimensional parameter, the electric Rayleigh number. Gradually increasing the electric Rayleigh number, the flow system sequentially evolves via quasi-periodic, periodic, and chaotic flow states with five identified bifurcations. The turbulence kinetic energy spectrum is shown to follow the -3 law as the flow approaches turbulence. The spectrum is found to follow a -5 law when the flow is periodic.

## I. INTRODUCTION

Electrohydrodynamics (EHD) is a multidisciplinary science that focuses on the interaction between fluid motion and the electric field[1]. Practical EHD applications attract extensive attention due to specific advantages, such as high efficiency, mobility, and controllability. In particular, EHD has been found promising in engineering applications, including propulsion[2, 3], particle removal[4, 5], heat transfer[6, 7], flow control[8, 9], and pumping[10, 11]. Despite the broad usage, a comprehensive stability analysis of the EHD flow remains to be unraveled because of the strong nonlinearity induced by the coupling effects between the flow and the electric field[12-14].

The electro-convection (EC) phenomenon is a fundamental focus of EHD research.

* Corresponding author. Tel.: +86 451 86402324; E-mail address: jian.wu@hit.edu.cn.

The Coulomb force acting on the free charge carriers destabilizes the system and drives the fluid flow, which in turn changes the charge distribution and the electric field due to the space charge effect. The unipolar injection mechanism is a simplification of the ion generation mechanism in dielectric liquids. It is assumed that the fluid is completely insulated and the unipolar charge injected from one electrode is the only source of free ions. In this work, the unipolar injection intensity remains constant during the injection process. This process can be induced by the complex electrochemical reaction at the electrode-fluid interface[15]. Taylor reported EC flows in 1966[16], describing cellular convection in the liquid droplet. Since then, EC flows in various geometric configurations have been investigated in many systems where the interaction of electrostatic force with fluids is present. Jolly et al.[17] demonstrated the EC flow with a unipolar charge injection experiment first in 1970. They found that the flow pattern changes with the increase of the electric Hartmann number. Atten et al.[18, 19] achieved stable and repeatable EC flows with unipolar injection by attaching an ion exchange membrane on the metal electrode surface and carrying out many experimental studies on dielectric liquid EC. In recent studies, the Laser Doppler Velocimetry (LDV) and Particle Image Velocimetry (PIV) techniques are widely applied to visualize the flow field[20-24].

With computer technology development, direct numerical simulation has become a widely accepted and effective method to gain additional insight into the EHD problems. To capture the large gradient and to avoid unphysical oscillations with high gradient electric charge density, various methods have been developed, such as the flux-corrected transport (FCT) schemes[25, 26] and the particle-in-cell (PIC) schemes[25, 26] of the finite-difference method (FDM), the total variation diminishing (TVD) schemes[27] of the finite-volume method (FVM), and the lattice Boltzmann method (LBM)[28-32]. In particular, the LBM has shown advantages such as inherent parallelism. Luo et al.[28-30] and Guan et al.[31, 32] applied an LBM to the EC flow and the electro-thermo-convection (ETC) flow.

Previous investigations of EHD have been primarily focused on steady-state laminar flows. Yet unsteady and turbulent convection is a fascinating topic in flow systems as they have ubiquitous practical applications. For a nonlinear hydrodynamic dissipative system, the oscillatory instability will cause a transition from the steady-state to a periodic oscillation when some critical condition is reached[33]. With a further increase in the parameter, the system will first undergo a sequence of instabilities resulting from the increasing nonlinearity, then gradually evolves toward chaos or weak turbulence, and eventually to fully developed turbulence[34]. Such a dynamic process is essential to both nonlinear hydrodynamics and industrial applications. Several classical transformation routes in mathematical models or simple systems have been verified by both theoretical and experimental methods, such as the Ruelle-Takens-Newhouse route through quasi-periodicity[35], the intermittency sequence[36], and the periodic-doubling sequence[37]. However, due to the complexity of the natural dynamic systems, many different new transition routes have been defined. For instance, an alternating sequence of periodic and chaotic regimes has been observed in both the mathematical model and experiments by Turner[38]. The transitions from periodic oscillatory state to quasi-periodic oscillatory state with frequency-locking phenomenon before the emergence of chaotic flow was observed in 2-D liquid layers of finite extent in Li's research[39]. Li and Su[30] investigated the transition process from laminar to chaotic flow in ETC flow of a dielectric liquid in a square cavity.

This work aims to provide a detailed description of flow characteristics during the transition from hydrostatic to chaotic flow states in EC flows of a planar dielectric liquid. This problem can be more complicated than the classical Rayleigh-Bérnard Convection (RBC) in terms of nonlinear dynamics. Firstly, due to the strong nonlinear coupling among the space charge distribution, the electric field, and the flow field, rich bifurcation and flow states can emerge. Secondly, a wide range of the control parameter, which depends on the applied voltage, can be realized experimentally. In this work, the two-relaxation-time lattice Boltzmann method (TRT LBM) approach is adopted[32] to study several major flow states and their bifurcation phenomena during the transition from the hydrostatic base state to the chaotic state with an increasing electric Rayleigh number.

The remainder of this paper is organized as follows: In Section II, the physical model and governing equations are presented and described. Section III presents the numerical method and boundary conditions for the study cases. A detailed discussion of the results is provided in Sec. IV. Concluding remarks are made in the last section.

## II. PHYSICAL MODEL AND GOVERNING EQUATIONS

Two-dimensional EC flow in an incompressible layer of Newtonian, dielectric liquid between two parallel planar electrodes is considered. The schematic of the flow configuration is presented in FIG. 1. The liquid is considered to be perfectly insulating with homogeneous and isotropic physical properties. The bottom electrode is connected to a Direct Current (DC) high voltage supply and is maintained at a high electric potential ($V_1$) while the top electrode is grounded ($V_0$, $V_0 < V_1$). Positive ions are injected from the lower electrode and collected by the upper electrode[31]. To simplify the discussion, the unipolar charge injection with density $q_0$ is assumed to be autonomous and homogeneous[40, 41]. The vertical distance between the electrodes is $H$, and a sampling point $A$ is chosen at the center of the computational domain to record the time series of physical variables.

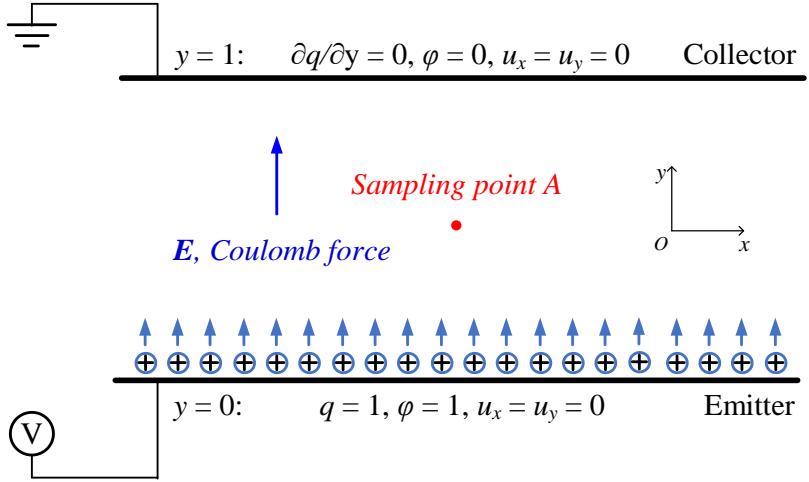

FIG. 1. Schematic diagram and non-dimensional boundary conditions of EHD convection in a layer of dielectric liquid.

The coupled set of governing equations include the Navier-Stokes (N-S) equations, the Poisson equation for electric potential, and the charge transport equation[42, 43]. The dimensional governing equations are given as follows:

$$\nabla \cdot \mathbf{u} = 0, \quad (1)$$

$$\rho \frac{D\mathbf{u}}{Dt} = -\nabla P + \eta \nabla^2 \mathbf{u} + q\mathbf{E}, \quad (2)$$

$$\frac{\partial q}{\partial t} + \nabla \cdot \left[ q(\mathbf{u} + K\mathbf{E}) - D_e \nabla q \right] = 0, \quad (3)$$

$$\nabla^2 \varphi = -\frac{q}{\varepsilon}, \quad \mathbf{E} = -\nabla \varphi \quad (4)$$

In the above equations, $\rho$ is the density of dielectric liquid; $\eta$ denotes the dynamic viscosity; $\mathbf{u} = (u_x, u_y)$ is the velocity vector; $P$ is the static pressure; $K$ is the ionic mobility; $D_e$ denotes the ionic diffusivity; $\varepsilon$ represents the dielectric permittivity; $\varphi$ represents the electric potential and $\mathbf{E} = (E_x, E_y)$ is the electric field. The term $q\mathbf{E}$ represents the Coulomb force.

Here, we consider $u_{drift} = K \Delta V / H$ as the characteristic velocity, where $\Delta V = V_1 - V_0$.

Other non-dimensional scales include $H$ for length, $H/u_{drift}$ for time, $\rho u_{drift}^2$ for pressure, and $\Delta V$ for $\varphi$. Following the above reference scales, the non-dimensional system of governing equations can be expressed as follows:

$$\nabla^* \cdot \mathbf{u}^* = 0, \tag{5}$$

$$\frac{D^* \mathbf{u}^*}{D^* t^*} = -\nabla^* P^* + \frac{M^2}{T}\nabla^{*2}\mathbf{u}^* + CM^2 q^* \mathbf{E}^*, \tag{6}$$

$$\frac{\partial^* q^*}{\partial^* t^*} + \nabla^* \cdot \left[ q^*\left(\mathbf{u}^* + \mathbf{E}^*\right) - \frac{1}{Fe}\nabla^* q^* \right] = 0, \tag{7}$$

$$\nabla^{*2}\varphi^* = -C q^*, \quad \mathbf{E}^* = -\nabla^* \varphi^* \tag{8}$$

These non-dimensional governing equations yield four following non-dimensional parameters:

$$C = \frac{q_0 H^2}{\varepsilon \Delta V}, \quad M = \frac{(\varepsilon/\rho)^{1/2}}{K}, \quad T = \frac{\varepsilon \Delta V}{\mu K}, \quad Fe = \frac{K \Delta V}{D_e} \tag{9}$$

The physical interpretations of these dimensionless parameters are as follows: $C$ represents the charge injection strength. $M$ is the ratio of the hydrodynamic mobility to ionic mobility. $T$ is the electric Rayleigh number, which represents the ratio of electric force to the viscous force, and $Fe$ is the reciprocal of the charge diffusivity coefficient[40, 43].

The non-dimensional boundary conditions are depicted in FIG. 1. The system is infinitely long in the horizontal direction, and therefore periodic boundary conditions are applied to all variables. The two electrodes are assumed to be impermeable, so the no-slip condition for velocity is used. The zero-gradient boundary condition specified for the charge density on the collection electrode means the charge does not accumulate on the planar electrodes but discharges after touching[44]. Note that the boundary conditions for charge density and electric potential used here are commonly used in EHD simulations[14, 31, 45].

## III. NUMERICAL METHODS AND CONDITION

In this paper, the TRT LBM approach[32] is applied to solve the transport equations for fluid flow and charge density, coupled to a fast Poisson solver for the electric potential. The numerical method is second-order accurate in space and time, and all variables are calculated in double precision to reduce the rounding errors. The macroscopic boundary conditions, mesoscopic boundary conditions are also specified in Table. 1, where $f_i(\mathbf{x},t)$ and $g_i(\mathbf{x},t)$ represent the discrete distribution function of velocity and charge density, respectively. The LBM full-way bounce-back (FBB) scheme is used for the Dirichlet (no-slip) boundary conditions for the fluid flow and charge density at the lower wall[29, 46, 47]. The $g_i$ Neumann boundary condition is set as a current outlet boundary condition for charge density transport[29, 46, 48]. More details about the TRT LBM approach and corresponding boundary conditions can be found in the recent publication[31, 32].

Table. 1 Boundary conditions for the numerical simulations

| Boundary | Macro-variables Conditions | Meso-variables Conditions |
|---|---|---|
| X-direction | Periodic | Periodic |
| Upper wall | $\mathbf{u}=0$, $\varphi=V_0$, $\partial q/\partial y=0$ | LBM FBB scheme for $f_i$ [48-52] <br> Neumann boundary condition $\partial g_i/\partial y = 0$ [48-52] |
| Lower wall | $\mathbf{u}=0$, $\varphi=V_1$, $q=q_0$ | LBM FBB for $f_i$ [48-52] <br> LBM FBB for $g_i$ [48-52] |

The numerical method is implemented in C++ using CUDA GPU computing. X-direction threads number in each GPU block is equal to the number of grids in X direction NX = LX/△ and the same as Y-direction, where LX and LY are proposed to represent the lengths in tangential and vertical directions, respectively. To improve computational efficiency while maintaining accuracy, a uniform grid of spacing △=0.005 is used throughout this work. In the present work, *LY* is set to be 1.0 to accompany the non-dimensional governing equations. To verify the tangential direction (X-direction) of the two-dimensional domain is appropriately sized for periodic conditions, the auto-correlation $R_{u'u'}(\Delta x) = \overline{u'(x,y,t)u'(x+\Delta x,y,t)} / \overline{u'(x,y,t)^2}$ at y = 0.5 is calculated[53], where $u'(x,y,t)$ is the fluctuating component of velocity $\mathbf{u}=(u_x,u_y)$ at time *t*. FIG .2 shows that the auto-correlation $R_{u'u'}$ is approaching 0 when △x > 1.23 which indicates that the domain (*LX*=2.46) is sufficiently large in the X-direction to avoid spurious coupling through the periodic boundary conditions.

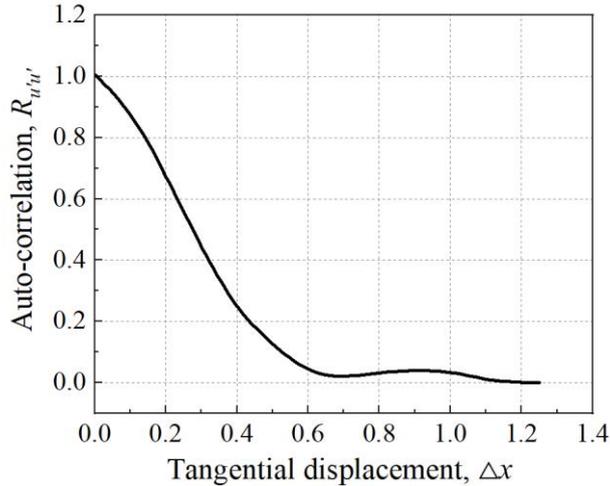

FIG. 2. Auto-correlation $R_{u'u'}$ of the fluctuating tangential velocity versus tangential displacement x at wall-normal positions y = 0.5 for $C$=10, $M$=10, $T$=2000, $Fe$=2000.

## IV. RESULTS AND DISCUSSION

The dielectric fluid layer develops from a hydrostatic base state to an unsteady state when the electric Rayleigh number $T$ is above a critical value. With a further increase of $T$, the flow exhibits different patterns and regimes before finally attaining the chaotic state. This section presents numerical results and statistical analyses of several distinguishable intermediate flow states during the transition to the chaotic flow state. The parameter $C$ is fixed at 10 to represent a strong injection strength, which can approximate the space-charge limited injection (SCL) condition[54]. The other non-dimensional parameters are $M$ = 10 and $Fe$ =2000, which are within reasonable ranges of dielectric liquids properties. These values are also investigated in previous numerical studies[31, 54]. Eight numerically identified flow states during the transition to chaos are summarized in Table. 2. The flow structure is qualitatively visualized by streamlines and charge density contours. Phase space diagrams are provided for the velocity fluctuation ($u' = u - \bar{u}$, where $\bar{u}$ is the time-average velocity) to demonstrate the states quantitatively. The Fourier analysis reveals the frequency spectra of the system.

To distinguish the quasi-periodic flow with multiple frequencies and the chaotic flow, nonlinear analysis methods including fractal dimension and Lyapunov exponent have also been adopted to further identify the emergence of chaos[55, 56]. The fractal dimension of the dynamical system counts the effective number of degrees of freedom and quantifies its complexity. The fractal dimension $d_c$ can be calculated by:

$$d_c = \lim_{s \to 0} \frac{\log(C_p(s))}{\log(s)} \tag{10}$$

$$C_p(s) = \lim_{N \to \infty} \frac{1}{N^2} \sum_{\substack{i,j=1 \\ i \neq j}}^{N} h(s - |X_i - X_j|) \tag{11}$$

where $N$ is the number of sampling points in p-dimensional space. The correlation function $C_p(s)$ represents the minimum number of hypercubes (of size s) covering all the points. $h$ is the Heaviside function. $|X_i - X_j|$ represents the distance between two given points in p-dimensional space. In particular, the fractal dimension of periodic flow is close to 1, that of quasi-periodic flow is close to 2, and that of chaos is greater than $2^{57}$. The maximum Lyapunov exponent $\lambda_{max}$ is an effective dynamic diagnosis method of chaotic systems. In the present work, the maximum Lyapunov exponent $\lambda_{max}$ of the charge density evolution at sampling point A is estimated by the algorithm of Wolf[58].

Table. 2 Flow characteristics summary of major EC flow states on the route from hydrostatic base state to chaos for $C=10$, $M=10$, $Fe=2000$: flow structure, fractal dimension, and maximum Lyapunov exponent with the increase of $T$.

| $T$ | Flow structure | Flow state | Fractional dimension | Maximum Lyapunov exponent |
|---|---|---|---|---|
| <155 | none | Rest | … | 0 |
| 160-300 | 4-cell | Steady | 0 | 0 |
| 310 | Multi-cell | Chaotic | 2.5 | 0.032 |
| 310-510 | Multi-cell | Chaotic | … | … |
| 520-550 | 4-cell | Quasi-periodic | 2 | 0 |
| 560-700 | 4-cell | Periodic | 1 | 0 |
| 710 | Multi-cell | Chaotic | 3.03 | 0.295 |
| >710 | Multi-cell | Chaotic | … | … |

**3.1 Transition from hydrostatic base state to steady EHD convection ($T=160 \sim 300$)**

The unipolar charge injection drives the dissipative flow system into motion, and the EC flow begins. When $T$ is greater than the critical value $T_1$ (155~160). FIG. 3-(a) and FIG. 3-(b) show the charge density evolution diagram and velocity phase space trajectory, respectively, at the sampling point A for $T=200$. The charge density reaches a constant value after a short fluctuation. The attractor morphology evolves into a stable fixed point, which indicates that the system eventually reaches a steady state. For a lower $T$, the system ultimately maintains a stable EC flow.

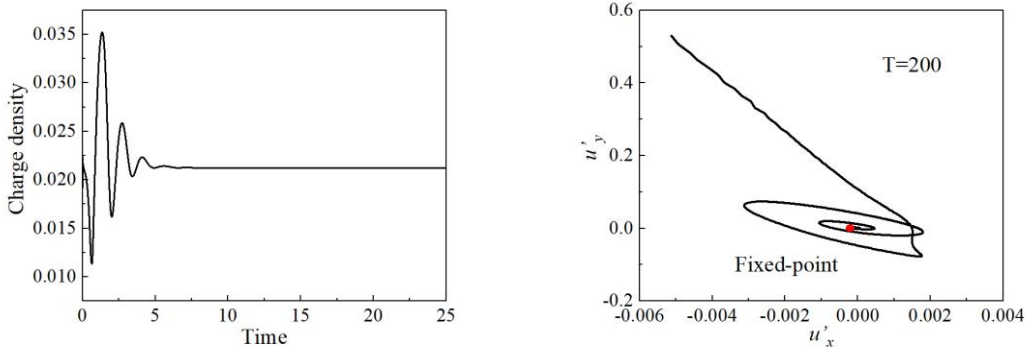

FIG. 3 Steady-state results at $T=200$: (a) Charge density evolution. (b) velocity phase space trajectory at sampling point A.

The charge density distribution with streamlines at $T=200$ is presented in FIG. 4. The flow pattern consists of two pairs of counter-rotating vortices. Two symmetric upward-moving plumes are noticed from the lower plate electrode. The charge density decays rapidly from bottom to top, forming a clear charge density gradient and two pairs of vortices of similar size and opposite directions.

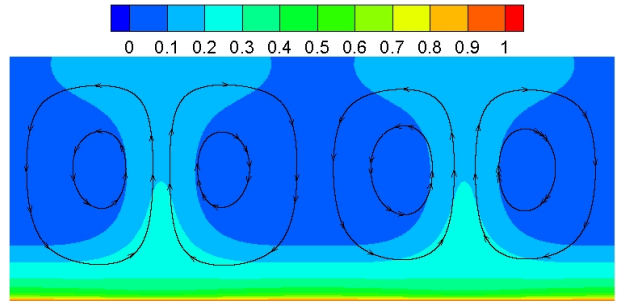

FIG. 4 Charge density distribution with streamlines at $T = 200$

## 3.2 Transition from steady convection to chaos ($T=310\sim510$)

When $T$ slightly exceeds the second bifurcation point $T_2$ between 300 and 310, new charge plumes are generated, and the flow becomes unsteady. The system loses its steady state and develops directly into a chaotic state. The fractal dimension $d_c$ is 2.5, and the maximum Lyapunov exponent $\lambda_{max}$ is 0.032 for charge density evolution at sampling point A when $T=310$, indicating the onset of a chaotic flow. The nonlinear characteristic of the system is visualized by the velocity phase space trajectories (left) and the corresponding Fourier spectra of charge density at sampling point A (right) for increasing $T$ (325-510) in FIG. 5. The disordered phase space trajectories and the typical continuous broadband characteristic illustrate the existence of chaos. The degree of phase space trajectory disorder first increases and then decreases with the increase of $T$. In particular, it is interesting that the phase space trajectories begin to form three limit cycles corresponding to three independent dominant frequencies in the Fourier spectrum for $T$=510. The appearance of limit cycles means that the system starts to have quasi-periodic characteristics. Therefore, $T$=510 indicates a critical value, and the dynamical system has both quasi-periodic and chaotic features.

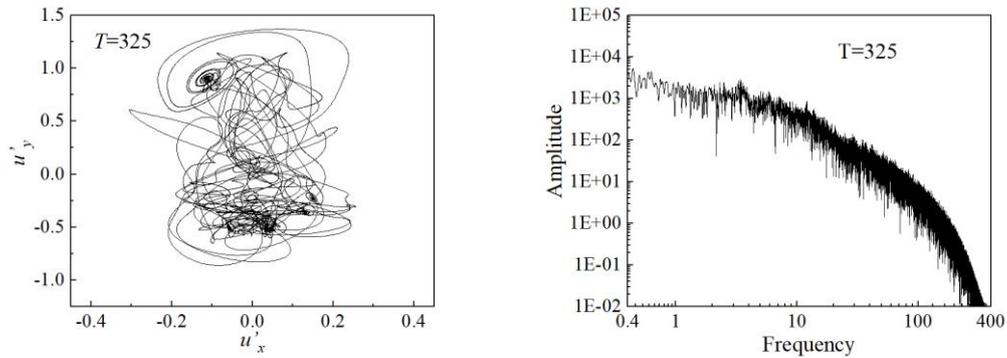

(a) $T$ =325, chaotic

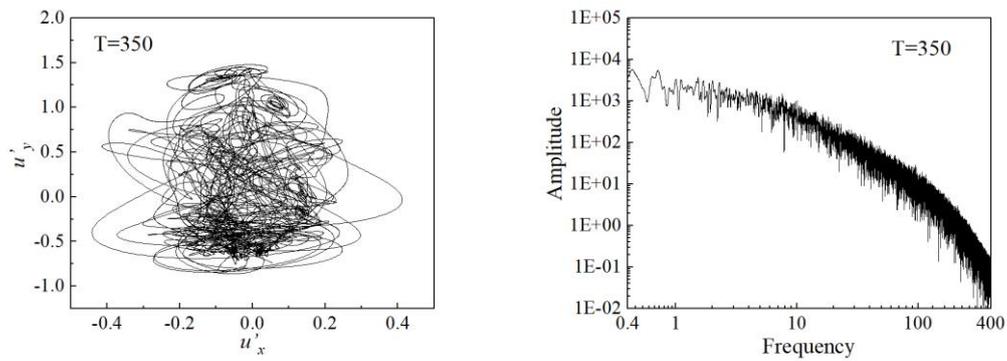

(b) $T$ =350, chaotic

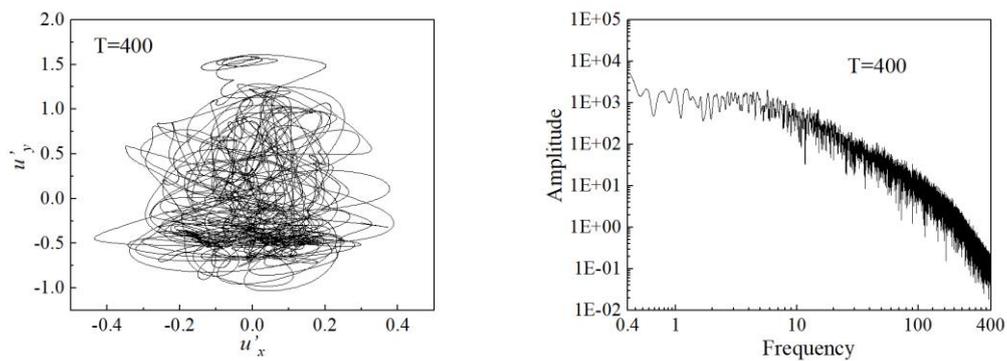

(c) T =400, chaotic

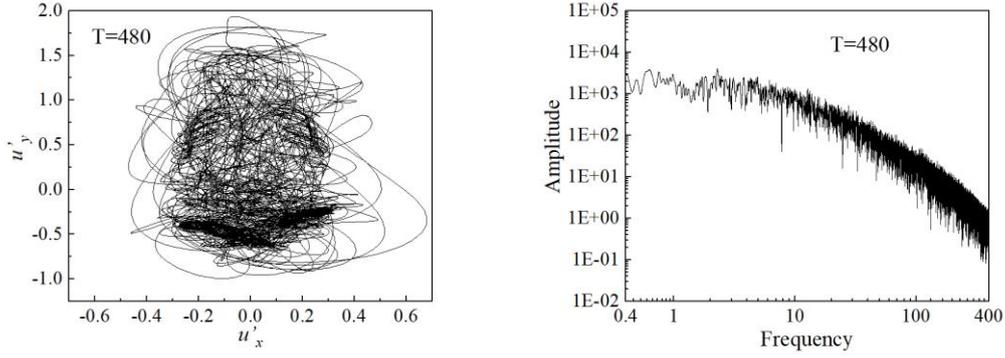

(d) T =480, chaotic

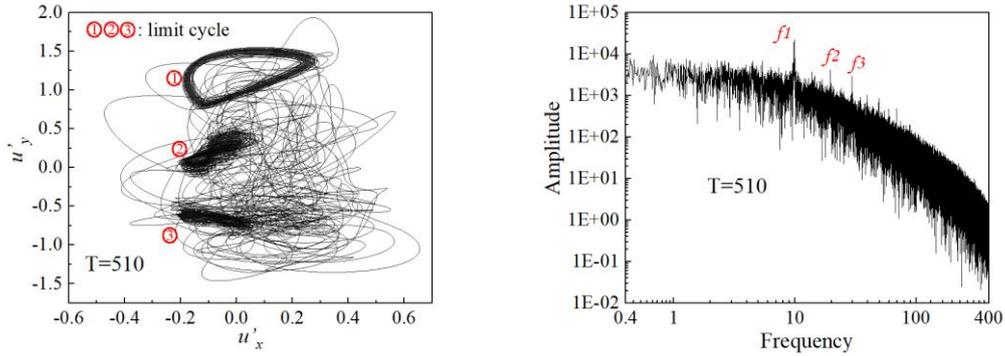

(e) T =510, chaotic

FIG. 5 Velocity phase space trajectories (left) and corresponding Fourier spectra of charge density at sampling point A (right) for increasing $T$: (a) $T$=325, chaotic, (b) $T$=350, chaotic, (c) $T$=400, chaotic, (d) $T$=480, chaotic, (e) $T$=510, chaotic.

The charge density evolution at the sampling point A for $T$ =310 is presented in FIG. 6-(a) and the intermittence characteristics demonstrated by a series of relatively regular large-amplitude bursting can be observed. To provide better insights into the intermittent flow, the ascending and descending parts are shown separately, and nine instantaneous snapshots of charge density distribution with streamlines corresponding to different time moments at $T$=310 marked in FIG. 6-(b) and FIG. 6-(c) are shown in FIG. 7. It can be seen that this intermittent phenomenon is mainly caused by the new plume arising between the two original plumes, and as the distance between the plumes is narrowed, plumes start to merge. The pattern changing begins with an increase of the oscillation amplitude (t1). The increased oscillation separates the two vortex pairs before a third plume gushes out in the middle of the domain bringing in an additional vortex pair. The three plumes are then squeezed into the middle of the domain (t3) and merge (t4) when another plume emerges at the periodic boundary (t5). The flow oscillation is small at t5, and the flow exhibits a nearly steady state. However, the oscillation amplitude gradually increases (t6), causing a new merging and emerging of plumes (t7, t8), after which the flow state returns to a shape similar to the one at t1 (t9).

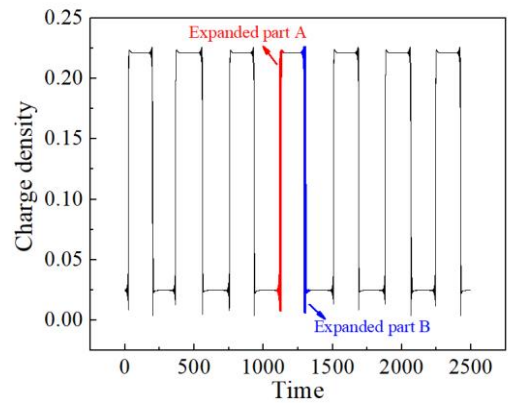

(a)

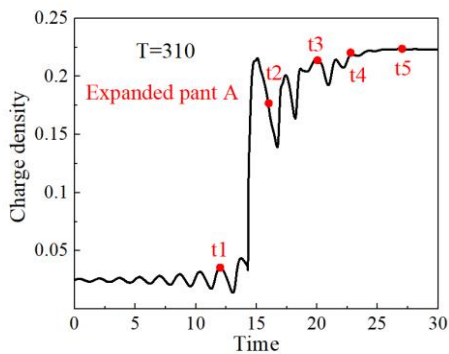 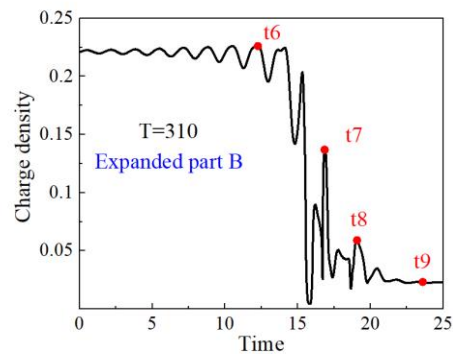

(b) (c)

FIG. 6 Evolution of the charge density at sampling point A for $T$ =310: (a) global charge density evolution, (b) the expanded ascending part A of the intermittent large-amplitude bursting, (c) the expanded descending part B of the intermittent large-amplitude bursting.

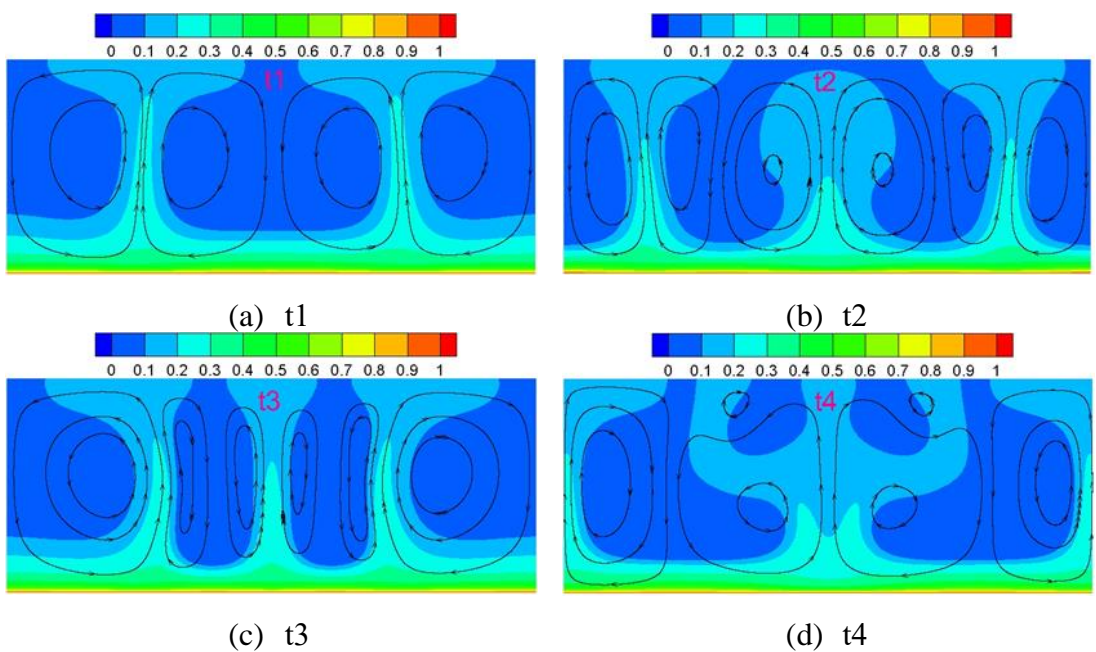

(a) t1  (b) t2

(c) t3  (d) t4

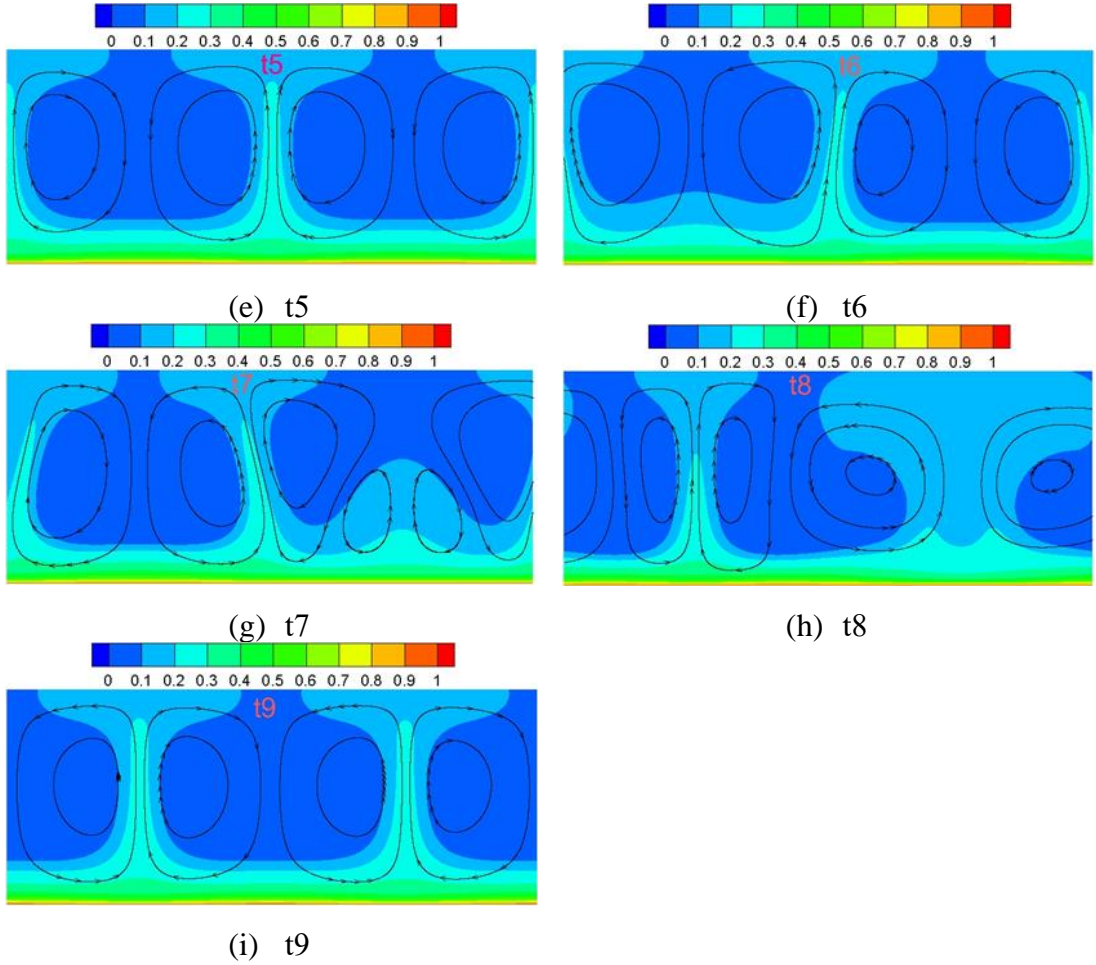

FIG. 7 Instantaneous snapshots of charge density distribution with streamlines corresponding to different time moments at $T=310$ marked in FIG. 6-(b) and FIG. 6-(c): (a)~(i) : t1~t9.

FIG. 8 compares the charge density evolution at sampling point for $T$=310, 315, and 325 within 500 non-dimensional units of time. The intermittent bursting has been significantly intensified with the increase of $T$, and the chaotic characteristic can be observed at $T$=325. The charge density evolutions at sampling point A for $T = 480$ and 510 are presented in FIG. 9. Fluctuations with smaller amplitudes can be observed, indicating the quasi-periodic characteristics of the system. The quasi-periodic cycles are more pronounced when $T$=510. It seems that $T$=510 is close to the critical bifurcation value between the quasi-periodic flow and chaos.

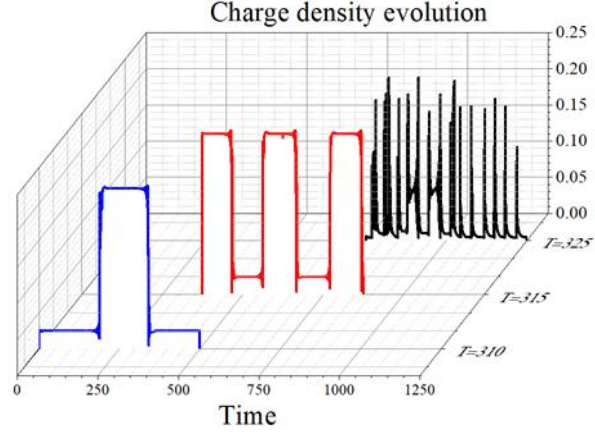

FIG. 8 Charge density evolution at sampling point A for $T$=310, 315, and 325 within 500 non-dimensional units of time

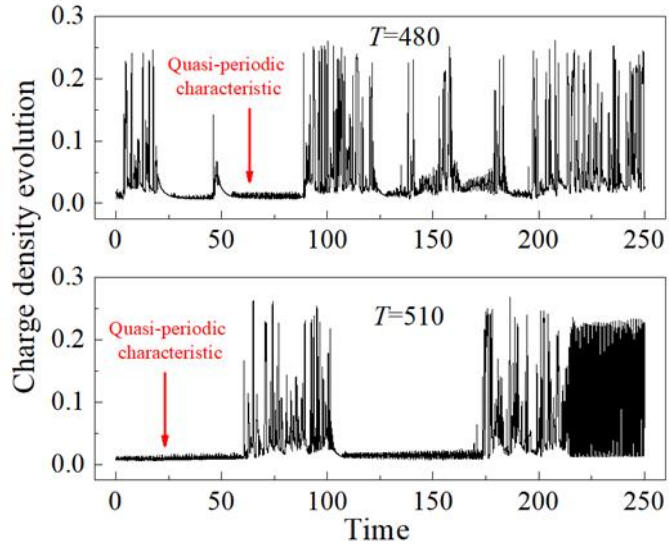

FIG. 9 Charge density evolution at sampling point A for T=480 and 510 within 250 non-dimensional units of time

**3.3 Transition from chaotic flow to Quasi-periodic ($T$=520~550)**

When $T$ exceeds the bifurcation $T_3$ between 510 to 520, the system evolves from chaos to a quasi-periodic flow. One plume exists in the center of the domain and swings left and right. With the increase of $T$, the fundamental frequency increases, and the quantity decreases. The maximum Lyapunov exponent $\lambda_{max}$ of the system decays to 0 for $T$=520, which indicates that the chaotic nature fades away. The fractal dimension $d_c$=2 proves the existence of a quasi-periodic flow cycle. The change in the coupling strength of attractors (quasi-periodic oscillators or chaotic oscillators) generated in the dissipative system is presumed to be responsible for this uncommon transition phenomenon[59, 60]. The work by Ontañón-García et al.[59] shows the bifurcation from chaos to quasi-periodic flow by changing the coupling strength of two bidirectionally linear interconnected Lorentz systems. Kuznetsov et al.[60] also proved a "self-oscillating component" in the dynamics of chaotic oscillators, which becomes apparent and initiates quasi-periodic oscillations when the dissipative coupling is introduced. The velocity phase space trajectories (left) and the corresponding Fourier spectra of charge density at the sampling point (right) for increasing $T$ (520-550) are presented in FIG. 10. For $T$=520, 540, and 560, the dominant frequency $f_1$ =9.96, 10.16, and 10.24, respectively. The quasi-periodic cycle frequency of the system increases with the increase of $T$. For $T$ <540, the Fourier spectrum consists of three incommensurable fundamental frequencies and all other peaks are harmonics, that have been confirmed to be a linear combination of the three main frequencies. A typical muti-tours limiting cycle surface is also presented in the phase space trajectories diagram. When $T$ is increased to 550, the number of fundamental frequencies are reduced to two. Therefore, we can expect that the number of fundamental frequencies will be reduced to 1 when the flow enters a periodic regime at the critical bifurcation value $T_4$, which will be shown in the next section.

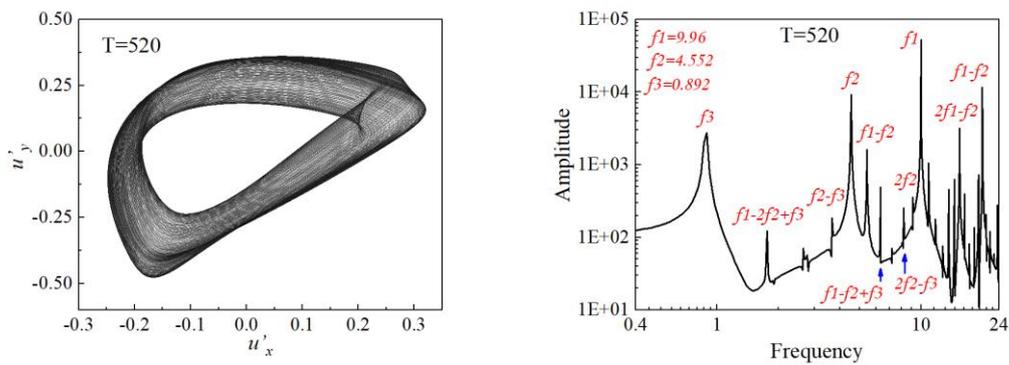

(a)  $T$ =520, quasi-periodic

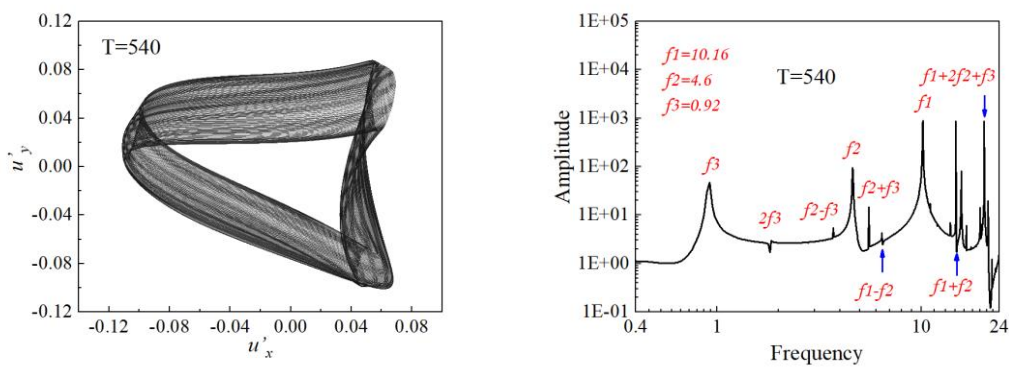

(b)  $T$ =540, quasi-periodic

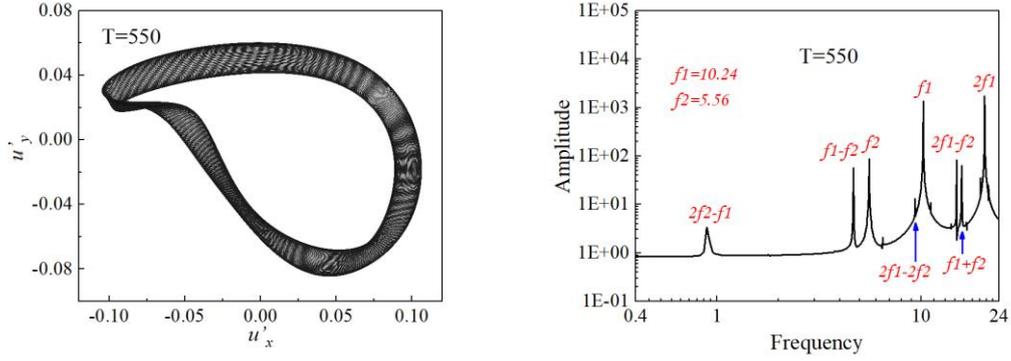

(c) $T=550$, quasi-periodic (critical state)

FIG. 10 Velocity phase space trajectories (left) and corresponding Fourier spectra of charge density at sampling point A (right) for increasing $T$: (a) $T=520$, quasi-periodic, (b) $T=540$, quasi-periodic, (c) $T=550$, quasi-periodic (critical state).

In FIG. 11, the charge density evolution at the sampling point for $T=520$ and two instantaneous snapshots of charge density distribution with streamlines corresponding to different time moments are shown in FIG. 12. The location and distribution of the charge plumes are very similar to the one shown in FIG. 7-t5. No more plumes gush out, and the charge plumes at the center of the domain swing horizontally leading to the squeezing and distortion of the flow vortices.

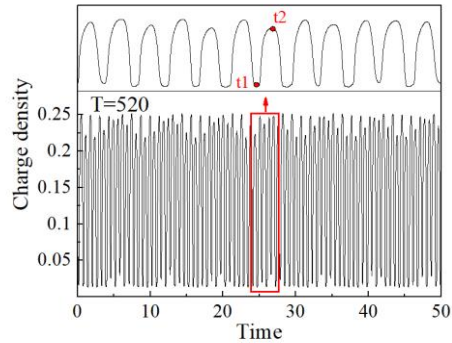

FIG. 11 Charge density evolution at sampling point A for $T=520$ within 50 non-dimensional units of time. The upper panel is the local part of the marked global evolution with the red rectangle.

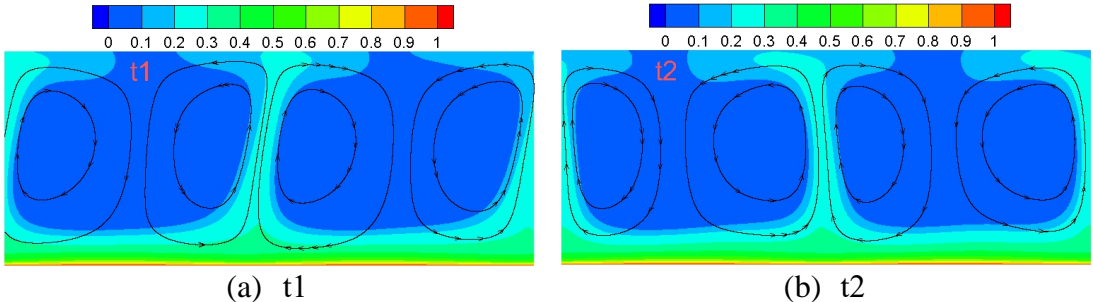

(a) t1                              (b) t2

FIG. 12 Instantaneous snapshots of charge density distribution with streamlines corresponding to different time moments at $T=520$ marked in FIG. 11: (a) t1, (b) t2.

## 3.4 Transition from Quasi-periodic flow to periodic flow ($T$=560~700)

A periodic (in time) flow regime begins when $T$ is greater than the fourth bifurcation critical value $T_4$. The velocity phase space trajectories (left) and corresponding Fourier spectra of charge density at the sampling point (right) for increasing $T$ (560-700) are shown in FIG. 13. The typical 1-tour phase space trajectories diagrams and the Fourier results with a single dominant frequency and an integer multiple of it confirm the periodic state. The dominant frequency $f_1$=10.304, 10.7, and 11.308 for $T$=560, 650, and 700, respectively.

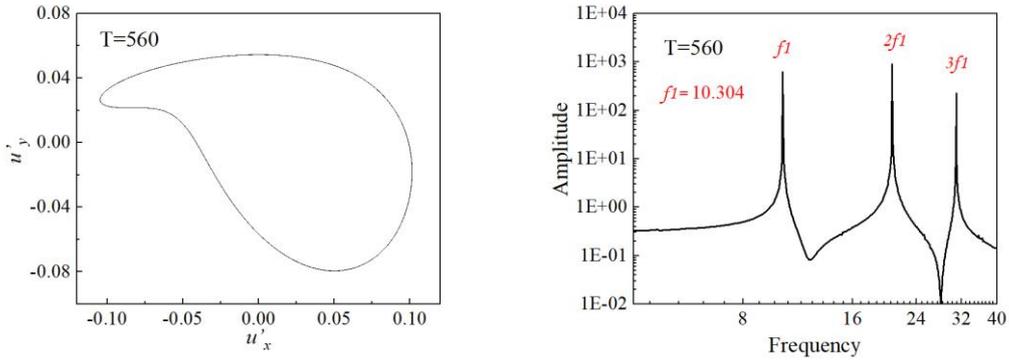

(a) $T$=560, periodic

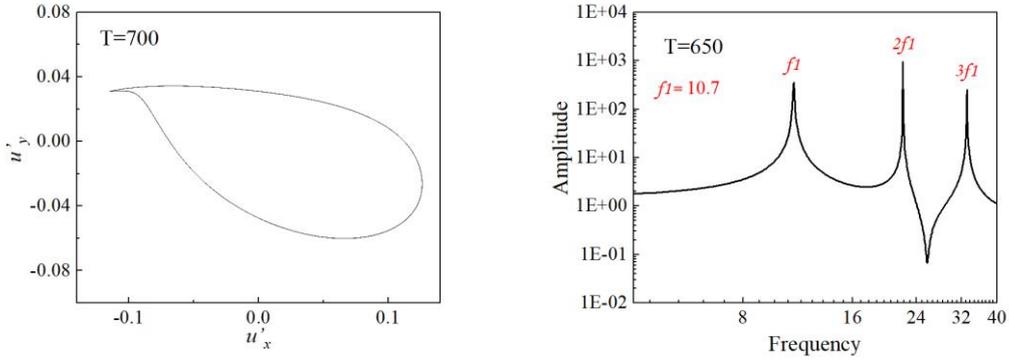

(b) $T$=650, periodic

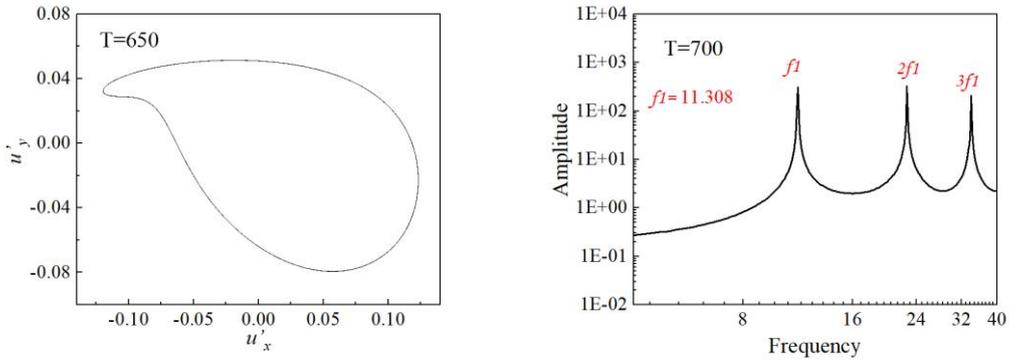

(c) $T$=700, periodic (critical state)

FIG. 13 Velocity phase space trajectories (left) and corresponding Fourier spectra of charge density at sampling point A (right) for increasing $T$: (a) $T$=560, periodic, (b) $T$=650, periodic, (c) $T$=700, periodic (critical state).

The charge density evolution at the sampling point for $T$=650 and two instantaneous snapshots of charge density distribution with streamlines corresponding to different time moments have been considered in FIG 14 and 15, respectively. The location distribution of the plumes is very similar to that in the steady-state ($T$=160~300), and the flow pattern is analogous to the quasi-periodic cycle ($T$=520~550). The two charge plumes emanating from the bottom electrode periodically swing in the transverse direction.

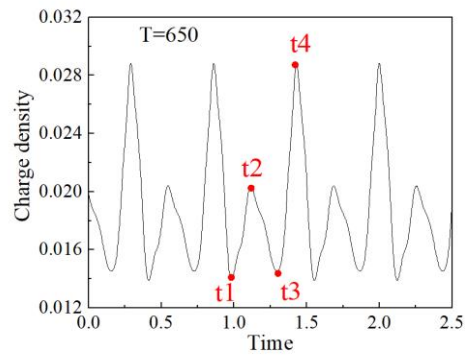

FIG. 14 Charge density evolution at sampling point A for $T$ =650 within 2.5 non-dimensional units of time

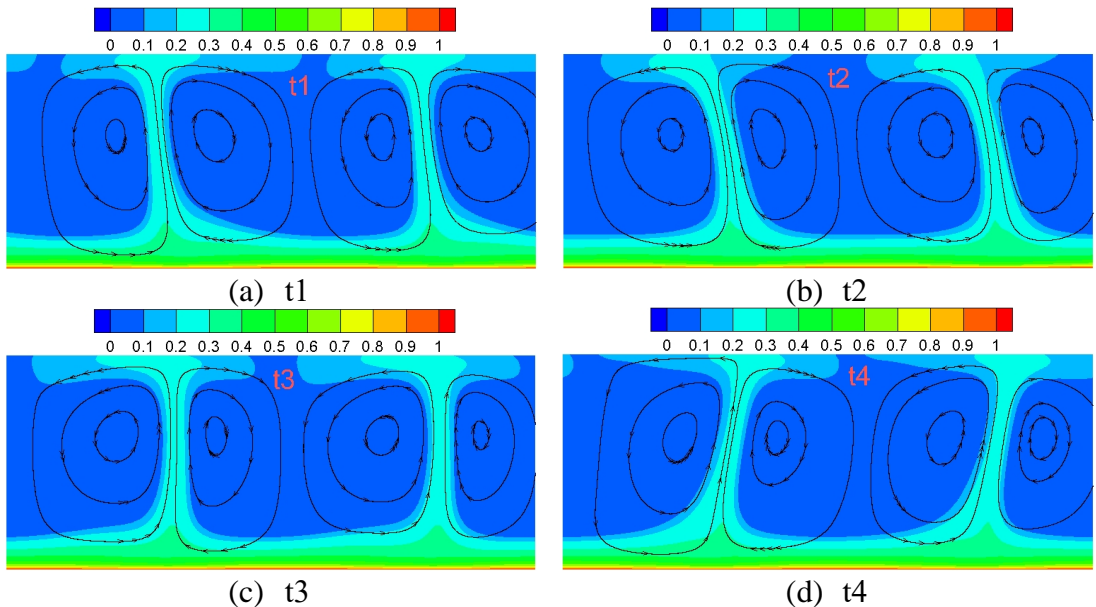

(a) t1    (b) t2

(c) t3    (d) t4

FIG. 15 Instantaneous snapshots of charge density distribution with streamlines corresponding to different time moments at $T$ =650 marked in FIG. 14: (a)~(d): t1~t4.

### 3.5 Transition from periodic flow to chaotic states ($T$>710)

With a further increase in $T$, the flow enters into a chaotic regime when $T$ exceeds the fifth bifurcation critical value $T_5$. The existence of chaos is confirmed by the fractal dimension $d_c$=3.03 and the maximum Lyapunov exponent $\lambda_{max}$=0.295 at $T$ = 710. FIG. 16 presents the velocity phase space trajectories (left) and corresponding Fourier spectra of charge density at the sampling point (right) for $T$=710, 800, and 900. The haphazard velocity phase space trajectories and the typical continuous broadband characteristic of the Fourier spectrum assure the existence of chaotic flow. With the increase in $T$, the phase space trajectory becomes more disordered, and the broadband characteristics of the Fourier result become more apparent, which means the system becomes more chaotic. The charge density evolution at the sampling point for $T$=800 and two instantaneous snapshots of charge density distribution with streamlines corresponding to different time moments are presented in FIG. 17 and 18. For chaotic systems at T=800, the horizontal fluctuating and shifting of plumes can be observed simultaneously. Comparing the result with the periodic cycle flow pattern, the swing horizontal fluctuation of plumes leads to oscillation. The plume's horizontal shifting is the reason for the dramatic bursting in charge density evolution for $T$ = 800 at the sampling point A as shown in FIG.17.

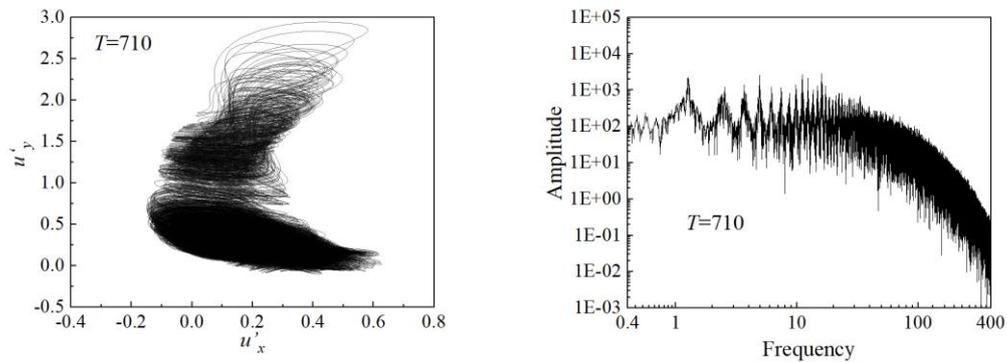

(a) $T$=710, chaotic

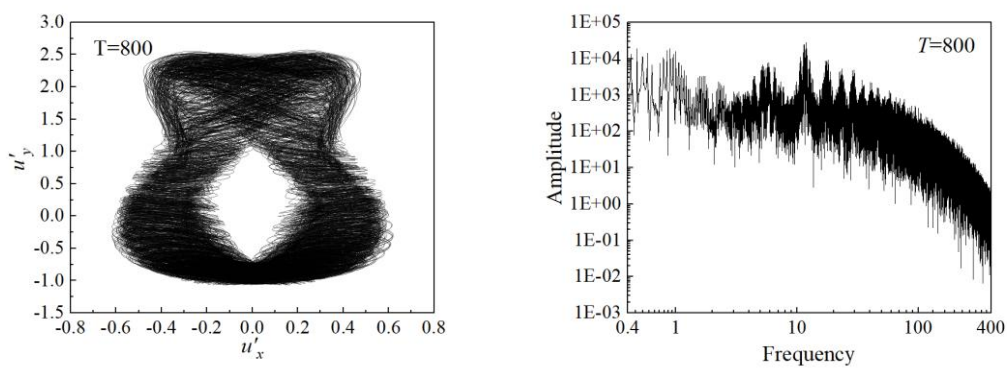

(b) $T$=800, chaotic

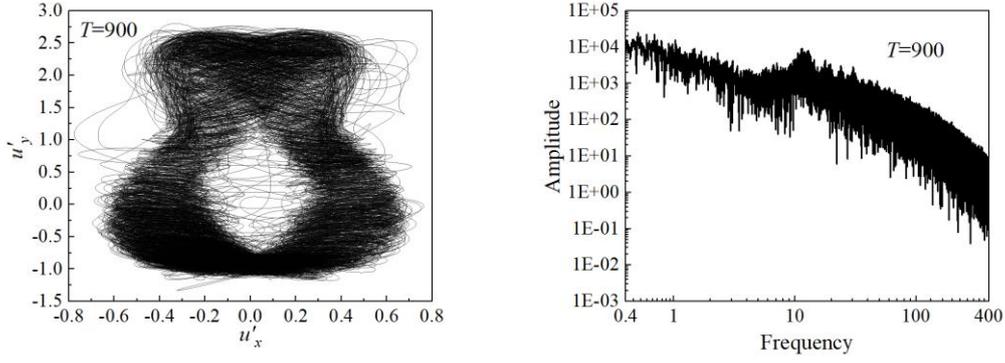

(c)  $T=900$, chaotic

FIG. 16 Velocity phase space trajectories (left) and corresponding Fourier spectra of charge density at sampling point A (right) for increasing $T$: (a) $T=710$, chaotic, (b) $T=800$, chaotic, (c) $T=900$, chaotic.

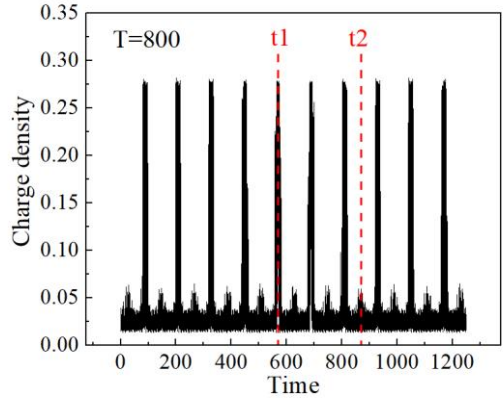

FIG. 17 Charge density evolution at sampling point A for $T=800$ within 1250 non-dimensional units of time

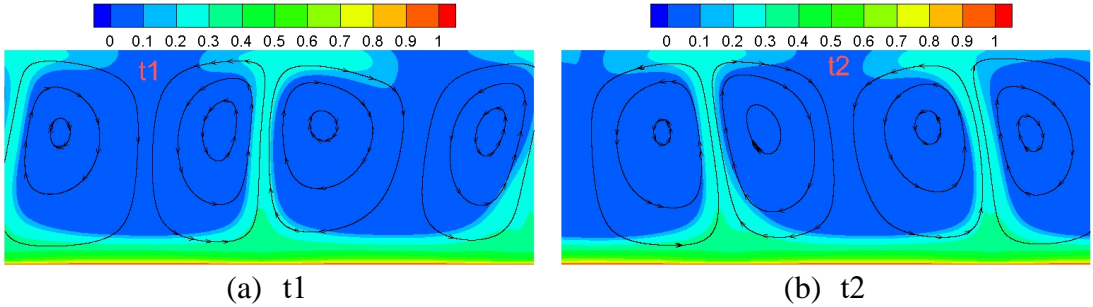

(a)  t1                              (b)  t2

FIG. 18 Instantaneous snapshots of charge density distribution with streamlines corresponding to different time moments at $T=800$ marked in FIG. 17: (a) t1, (b) t2.

### 3.6 Identified sequential flow states and the power law of turbulent kinetic energy

In the above sections, the nonlinear characteristics of the EC flow states are determined with a flow of increasing $T$. The major flow states on the transition route from the hydrostatic base state to the chaos of the EC system between two horizontal parallel planar electrodes are identified. The major flow states with five bifurcation points are summarized in Fig19. We observe intermediate periodic/quasi-periodic states on the route to chaos with an increasing control parameter ($T$ in our case), which is uncommon in dynamical systems.

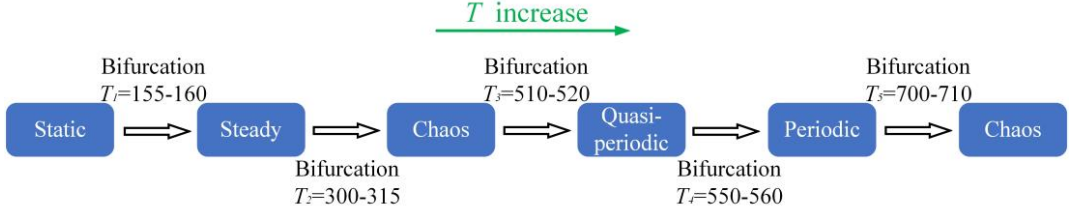

FIG. 19 The major flow states with five bifurcation points on the transition route from the hydrostatic base state to the chaos of the EC system of a dielectric liquid between two parallel planar electrodes for $C=10$, $M=10$, $Fe=2000$ with a flow of increasing $T$.

FIG. 20 shows the ensembled averaged turbulent kinetic energy (TKE) spectra at various y locations of four flow states near the bifurcation points. The TKE spectra are shown to follow the classical 2D turbulence power law (TKE ~ $k^{-3}$) at the inertial subrange for $T = 800$ due to its proximity to turbulence. At the quasi-periodic or periodic flow states, however, the TKE spectra follows $k^{-5}$. For the critical value $T=500$, the slope is between -3 and -5 due to the coexistence of periodic and chaotic features.

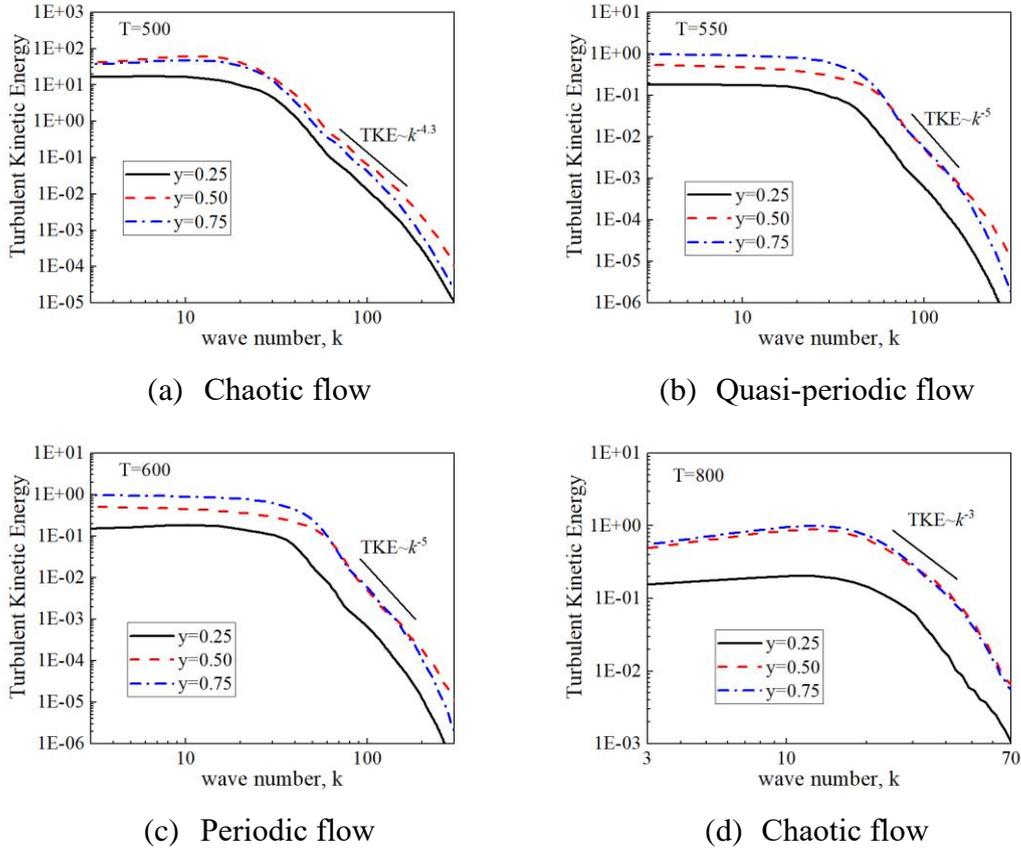

(a) Chaotic flow  (b) Quasi-periodic flow

(c) Periodic flow  (d) Chaotic flow

FIG. 20 Spatial spectra of turbulence kinetic energy at different y loaction: (a) $T=500$, (b) $T=550$, (c) $T=600$, (d) $T=800$.

## 3.7 The influence of *C* and *M*

The transition to chaos when *C* and *M* are fixed is presented above with an increasing electric Rayleigh number *T*. In this section, a qualitative analysis of the influence of *C* (injection strength) and *M* (mobility) on the transition behavior is discussed.

In non-dimensional EHD systems, *C* represents the charge injection strength. Generally, *C*=1 is considered to be the "medium" injection[61]. *M* is the ratio of hydrodynamic mobility to ionic mobility, which can represent the capacity of charge transport in a dielectric liquid. The major flow states on the transition route to chaotic flow states for *C*=1, *M*=10, and *C*=10, *M*=20 are identified in FIG. 21.

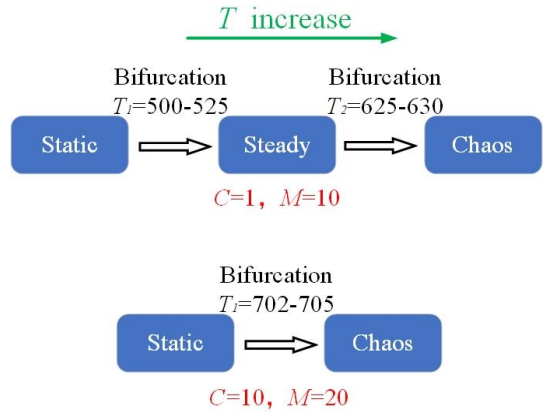

FIG. 21 The major flow states with bifurcation points on the transition route from the hydrostatic base state to chaos of the EC system of a dielectric liquid between two parallel planar electrodes for *C*=1, *M*=10, *Fe*=2000 and *C*=10, *M*=20, *Fe*=2000.

With the increase of electric Rayleigh number *T*, the unipolar charge injection drives the dissipative flow system into motion. For *C*=1, when *T* is greater than the first critical value (*T1*=525), the EC flow begins, and when *T*>650 (*T2*), the system becomes chaotic. Comparing to FIG. 19, the decrease of *C* increases the critical value of *T1* and reduces the number of bifurcation points.

For *M*=20, the convective flow occurs when *T* exceeds the critical value (*T1*=702). At this *T* value, the fractal dimension $d_c$ of the flow state is 2.5, and the maximum Lyapunov exponent $\lambda_{max}$ is 0.26, indicating that the system becomes chaotic, and no intermediate states are captured in our numerical simulation (with a parameter accuracy of Δ*T*=1). The direct transition from the hydrostatic base state to chaos by an increasing *T* at large *M* values has also been observed in the EC flow with weak injection conditions (*C*=0.1)[62, 63]. The charge density evolution at sampling point A is shown in FIG .22, and a series of large-amplitude bursting can be observed at *T*=702.

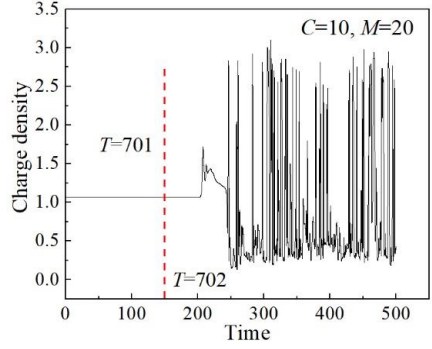

FIG. 22 Evolution of the charge density at sampling point A for *T*=701 (hydrostatic state) and 702 (chaos) when *C*=10, *M*=20.

The nonlinear characteristic is visualized by the velocity phase space trajectories (left) and the corresponding Fourier spectra of charge density at sampling point A (right) at *T*=702 in FIG. 23. The disordered phase space trajectories and the typical continuous broadband characteristic also indicate the existence of chaos.

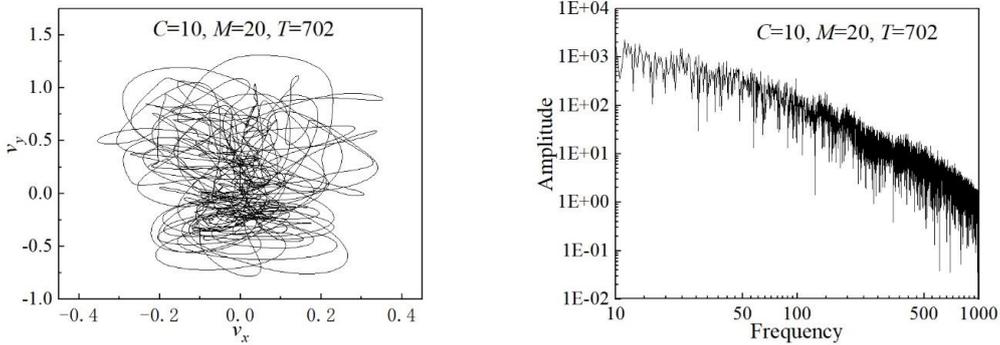

FIG. 23 Velocity phase space trajectories (left) and corresponding Fourier spectra of charge density at sampling point A (right) for *T*=702.

Comparing to FIG. 19, the increase of the mobility parameter *M* leads to a direct transition to chaos with fewer intermediate states. However, there is currently no widely accepted physical mechanism that can explain this behavior. We notice that $M^2$ in Eq. (6) plays both the stabilizing role in terms of viscous term and the destabilizing role in the driving force. However, along with the increase of *T*, the viscosity coefficient decreases, the driving force becomes relatively more robust. The inherent physical reasons can be the future research focus.

## IV. CONCLUSION

The major flow states on the transition route from hydrostatic base state to chaos in an EC flow system of a dielectric liquid between two infinite parallel planar electrodes are investigated via a TRT-LBM approach. The computational model is implemented in C++ using CUDA GPU computing. With an increasing electric Rayleigh number $T$, the system exhibits a variety of intermediate flow states such as the quasi-periodic, periodic, and chaotic states with different dynamical behaviors and five identified bifurcation points. When $T$ slightly exceeds the first critical value ($T1$), the system exhibits a steady-state flow followed by a chaotic flow with intermittent bursting behavior when $T>T2$. With a further increase of $T$, the dissipative system evolves from chaos to a quasi-periodic state, which is followed by a periodic flow when $T>T4$. During this process, the number of fundamental frequencies decreases while the dominant frequency increases. At a higher $T$ value, the system exhibits a transition from periodic flow to chaotic states, and the horizontal movement of the charge plumes are observed. The power law of turbulence kinetic energy (TKE) spectra changes at each bifurcation point. The typical 2D turbulence power law at $k^{-3}$ can be adopted for moderate EHD turbulence. The spectra are shown to follow a $k^{-5}$ law when a quasi-periodic or periodic flow is observed. For critical $T=500$, the slope is between -3 and -5 due to the coexistence of periodic and chaotic characteristics. The influence of the variation of the injection strength $C$ and the mobility ratio $M$ is also qualitatively discussed. It is found that the decrease of $C$ and the increase of $M$ both reduce the number of bifurcation points. This work enriches the research scope of nonlinear dynamics, and the analysis methodologies can be generalized to other convective flows such as RBC and magneto-hydrodynamic convection. To further explore the EC flow between two parallel electrodes, the transition route to turbulence in 3D EHD flow can be a future research topic.

## V. DATA AVAILABILITY

The data that support the findings of this study are available from the corresponding author upon reasonable request.

## VI. ACKNOWLEDGEMENTS

This work is supported by the National Natural Science Foundation of China (Grant No. 11802079).